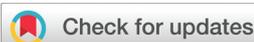

Check for updates

# Lattice induced crystallization of nanodroplets: the role of finite-size effects and substrate properties in controlling polymorphism

Julien Lam * and James F. Lutsko

Targeting specific technological applications requires the control of nanoparticle properties, especially the crystalline polymorph. Freezing a nanodroplet deposited on a solid substrate leads to the formation of crystalline structures. We study the inherent mechanisms underlying this general phenomenon by means of molecular dynamics simulations. Our work shows that different crystal structures can be selected by finely tuning the solid substrate lattice parameter. Indeed, while for our system, face-centered cubic is usually the most preponderant structure, the growth of two distinct polymorphs, hexagonal centered packing and body-centered cubic, was also observed even when the solid substrate was face-centered cubic. Finally, we also demonstrated that the growth of hexagonal centered packing is conditioned by the appearance of large enough body-centered cubic clusters thus suggesting the presence of a cross-nucleation pathway. Our results provide insights into the impact of nanoscale effects and solid substrate properties towards the growth of polymorphic nanomaterials.

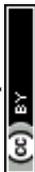

## 1 Introduction

When a nanoscopic liquid droplet is deposited onto a solid surface and rapidly cooled down, heterogeneous crystallization is triggered and nanocrystals are synthesized. This method has attracted much attention thanks to its rapid implementation and flexibility, for example, in pulsed laser deposition,[1–3] sputter deposition[4–6] and plasma-enhanced chemical vapor deposition.[7,8] The underlying mechanisms have been recently investigated experimentally and nanocrystal growth was directly observed for particular inorganic materials using high-resolution transmission microscopy[9–11] and X-ray scattering.[12] However, a major challenge in physics and chemistry remains the relationship between the synthesis conditions and the final outcomes especially in terms of crystal structure. Polymorphism is observed in numerous materials among which are protein crystals,[13–15] inorganic solids[16–19] and colloidal self-assembly.[20–22] It plays an important role in materials science since the crystalline structure determines the physical properties of the material and a fortiori its technological applications. At the nanometric scale, polymorphism is enhanced by the preponderance of surface effects promoting novel structures that would be unstable otherwise.[23–28]

In lattice induced crystallization of nanodroplets, the complexity results from two specific features. On the one hand, previous studies showed how finite-size effects can affect the wetting properties of liquid nanodroplets.[29–32] Nonclassical nucleation mechanisms are also observed when decreasing sizes down to the nanometric scale.[33] On the other hand, numerous studies have investigated the role of the substrate in enhancing bulk crystallization using model systems such as hard-sphere[34–37] and Lennard-Jones.[38] In particular, it was shown that the nucleation rate peaks when the lattice substrate does not perfectly match that of the nucleating crystal.[38] For a more complex system such as ice nucleation, Fitzner et al. demonstrated that the morphology of the lattice substrate can even lead to the formation of different ice polymorphs.[39]

However, the interplay between (a) the finite-size of the nanoscopic droplet and (b) the substrate properties which are characterized by its hydrophobicity and its lattice structure has received considerably less attention from the modeling point of view.[40] Consequently, a qualitative picture relating these two fundamental features to the crystal polymorphism is still far from being achieved. In this article, we used the molecular dynamics simulations of Lennard-Jones particles to follow the freezing of a nanoscopic liquid droplet deposited onto a crystalline substrate. We studied the effect of the lattice parameter and of the droplet size while maintaining the substrate crystalline structure to be face-centered cubic (FCC). Surprisingly, even if FCC is expected to arise, body-centered cubic (BCC) plays a critical role by emerging as

Center for Nonlinear Phenomena and Complex Systems, Code Postal 231,
Université Libre de Bruxelles, Boulevard du Triomphe, 1050 Brussels, Belgium.
E-mail: julien.lam@ulb.ac.be

the preponderant structure under certain conditions and by being a nucleation intermediate triggering the growth of hexagonal compact packing (HCP) under other conditions. The model is designed to be simple in order to capture the main physical features of this challenging problem and draw general conclusions on how to control the nanocrystal polymorphism.

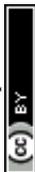

## 2 Methods

The system is composed of two types of particles: (1) particles that are kept fixed to represent the solid substrate designated as s and (2) particles that are moving designated as m. The system temperature denoted as $T$ was kept fixed using a Nosé–Hoover thermostat. All of the particles interact with a truncated and shifted Lennard-Jones (LJ) potential. For the moving particles, the LJ energy and length scale parameters are respectively denoted as $\varepsilon$ and $\sigma$. The cutoff was chosen equal to $3\sigma$. To initialize the simulations, a spherical-cap droplet of liquid is deposited onto the substrate with a lower density gas occupying the remaining simulation volume. The liquid and gas densities were chosen according to the conditions of liquid/gas coexistence at $k_B T = 0.8\varepsilon$ which gives $0.761\sigma^{-3}$ for the liquid and $0.01041\sigma^{-3}$ for the gas.[41] The system was first run at $k_B T = 0.8\varepsilon$ during 1000 MD steps to allow for some relaxation of the fluid. The initial droplet shape is not necessarily in equilibrium especially for the most hydrophilic substrate. Yet, it corresponds to experimental conditions where the droplets are deposited and directly quenched in temperature. The investigated substrate was represented by particles disposed on an FCC lattice with a (100) exposed plane. The zero-temperature FCC equilibrium density for Lennard-Jones is $\rho_0 = 1.09\sigma^{-3}$ (ref. 42) which corresponds to a lattice parameter equal to $a_0 = (4/\rho_0)^{1/3}$. The substrate lattice parameter is denoted as $a_s$ and the substrate particles were kept fixed.[43] Finally, wall substrate particles interact with the moving particles with $\sigma_{s/m} = \sigma$ and with LJ energy parameter $\varepsilon_{s/m}$. Even if long interactions such as electrostatic ones are not present in this model, hydrophobicity can be well tuned simply by changing the ratio $\varepsilon_{s/m}/\varepsilon$.[31,44] $N$ denotes the initial number of particles in the liquid droplet. Periodic boundary conditions were applied. The simulation boxes were chosen commensurate with the substrate lattice parameter. In addition, self-interactions were avoided by using simulation boxes larger than three (two) times the initial liquid droplet diameter in the planar (normal) directions. The simulations were carried out using LAMMPS.[45] Time is reported as a multiple of $t_0 = 500$ MD time steps.

Crystalline structures were identified by using an adaptive common neighbor analysis (CNA)[46,47] as implemented in OVITO[48] [see Fig. 1]. In this work, four independent parameters are varied: (1) the temperature $T$, (2) the initial liquid droplet size $N$, (3) the lattice parameter of the substrate $a_s$ and (4) the interaction energy between the substrate and the moving particles $\varepsilon_{s/m}$.

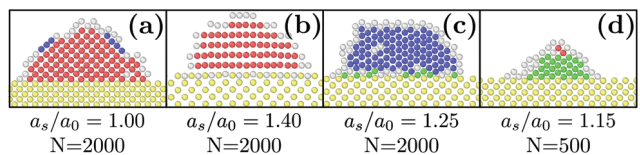

**Fig. 1** Slices of typical final states obtained with $\varepsilon_{s/m} = 2\varepsilon$, $k_B T = 0.3\varepsilon$ and different values of $N$ and $a_s/a_0$. Yellow particles represent the substrate. Red, blue, green and white colors designate respectively FCC, HCP, BCC and liquid particles.

## 3 Results and discussion

First, we investigated the influence of the substrate wetting properties. In general, the substrate is hydrophilic when it attracts more the moving particles than two moving particles attract themselves. For this study, the substrate lattice parameter is chosen to match the bulk non-deformed crystal ($a_s = 1.0 a_0$). As shown in Fig. 1a, the only emerging crystalline structure is FCC since the substrate lattice is not strained. Fig. 2a shows that in this case, the number of FCC particles grows and finally reaches a plateau. Then, averaging over 10 independent runs, we computed the value of the plateau, $N_{FCC}^\infty$, and the time at which half the plateau is reached, $\tau_{1/2}$ [see Fig. 2b]. Two regimes are observed: (i) when $\varepsilon_{s/m} > 1\varepsilon$, the substrate is so hydrophilic that increasing $\varepsilon_{s/m}$ has little influence over both $\tau_{1/2}$ and $N_{FCC}^\infty$ and (ii) when $\varepsilon_{s/m} < 1\varepsilon$, the substrate becomes more hydrophobic which suppresses heterogeneous crystallization as seen in the increase of $\tau_{1/2}$ and the decrease

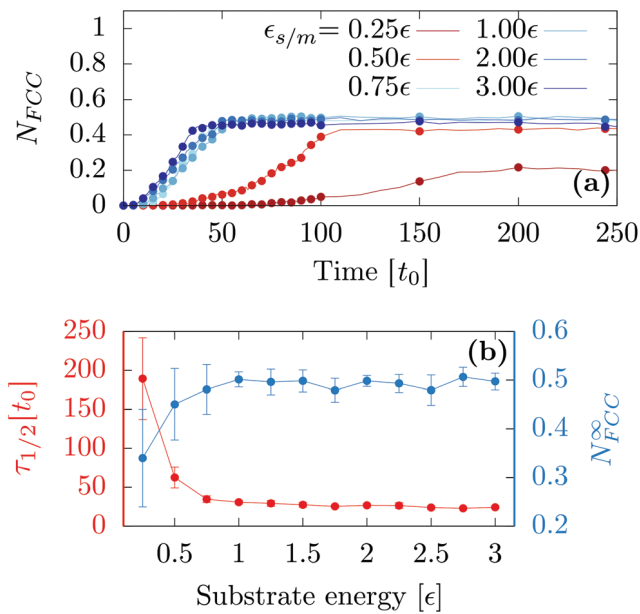

**Fig. 2** (a) Proportion of FCC particles as a function of time for a typical MD trajectory. (b) Average first passage time $\tau_{1/2}$ (on the left) and the average number of final FCC particles (on the right) as a function of the energy. Calculations are carried out with $k_B T = 0.3\varepsilon$, $N = 2 \times 10^3$ and $a_s = 1.0 a_0$.

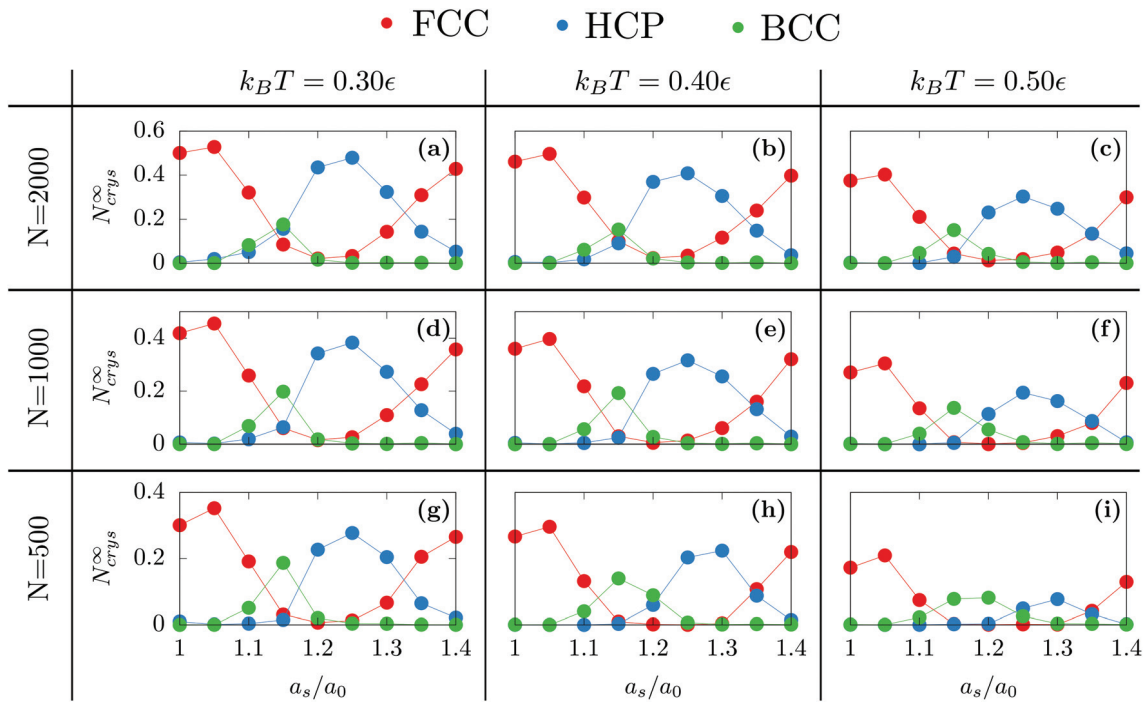

**Fig. 3** Proportion of crystalline particles denoted as $N^\infty_{crys}$ as a function of the rescaled substrate lattice parameter $a_s/a_0$ obtained with $\varepsilon_{s/m} = 2.00$ for different temperatures and nanodroplet sizes.

of $N^\infty_{FCC}$. For the rest of the article, we will focus solely on the case of a very hydrophilic substrate $\varepsilon_{s/m} = 2\varepsilon$.

Following the same procedure, we can compute $N^\infty_{crys}$ for FCC, HCP and BCC when varying the temperature, the nanodroplet size and the lattice parameter $a_s$ [see Fig. 3]. When considering the overall crystallinity, two general features can be observed. On the one hand, decreasing the droplet size leads to a decrease of the overall crystallinity. This is because there are more and more particles located at the droplet surface and these do not contribute to the overall crystallinity [see Fig. 1]. On the other hand, varying the temperature affects the quality of crystallization with lower temperatures yielding higher overall crystallinity. Finally and most interestingly, while the growth of hexagonal structures on top of an hexagonal substrate was previously obtained,[40] we will show that other crystalline structures can also be obtained when changing the substrate lattice parameter.

### 3.1 Different polymorphs

When $a_s \leq 1.10a_0$, the substrate is so similar to the bulk non-deformed crystal that FCC is the most preponderant structure [see Fig. 1a]. This result also holds when $a_s = 1.40a_0$. In this case, $a_s$ matches the second nearest neighbor distance of the bulk non-deformed crystal thus allowing to form the FCC crystal but with a different orientation [see Fig. 1b]. The highest crystallinity is obtained at $a_s = 1.05a_0$ which is consistent with the results obtained by Mithen and Sear who showed that the nucleation rate peaks for slightly stretched crystalline substrate.[38]

At the intermediate value of $a_s = 1.25a_0$, HCP becomes the most preponderant structure. In general, HCP is entropically less favorable than FCC so that it is only the equilibrium state at low enough temperature.[49,50] While this may explain the results obtained at $k_BT = 0.3$, for the larger temperature, the preponderance of HCP can only result from the combined presence of the substrate and the nanometric scale.

Finally, as evidenced in Fig. 1d and 3, the spontaneous emergence of BCC at $a_s = 1.15a_0$ is quite surprising since the bulk phase diagram of the Lennard-Jones crystal does not exhibit any BCC regions.[50] Yet, it is another confirmation that the substrate as used here is a way to stabilize exotic structures. In addition, when increasing the droplet size, BCC particles become less preponderant thus highlighting the decisive role of nanometric size effects.

### 3.2 Role of the first layer

In order to understand these striking results, we investigated the role of the first growing layer using Grand Canonical Monte Carlo run at $T = 0.3\varepsilon$. The chemical potential is chosen to match the binding energy of the FCC Lennard-Jones crystal at $T = 0.3\varepsilon$ which is $\mu = -6.50\varepsilon$.[51] For a given strain $a_s$, the initial structure is a strained crystalline substrate with periodic boundaries on the two planar axes and a width of $2a_s$. In the normal axis, we imposed a non-periodic and fixed boundary. The box size was $10a_s \times 10a_s \times 3a_s$ thus leaving $1a_s$ width to be filled with moving particles. This approach allows us to probe the first growing layer without any bias concerning its initial structure. Fig. 4 shows the resulting structures. As in the pre-vious simulations on droplet freezing, FCC structures are observed for $a_s \leq 1.10a_0$ and for $a_s \geq 1.35a_0$ thus confirming that these values of the strain favor the growth of FCC in the droplet configuration. Similarly, for $a_s = 1.15a_0$, the first growing layer is made of BCC structures as in the droplet configuration. However, in between, for $1.20a_0 \leq a_s \leq 1.25a_0$, the first growing layer is also made of BCC structures while HCP droplets are formed. In the next section, we will discuss the role of this BCC layer in the growth of HCP.

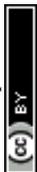

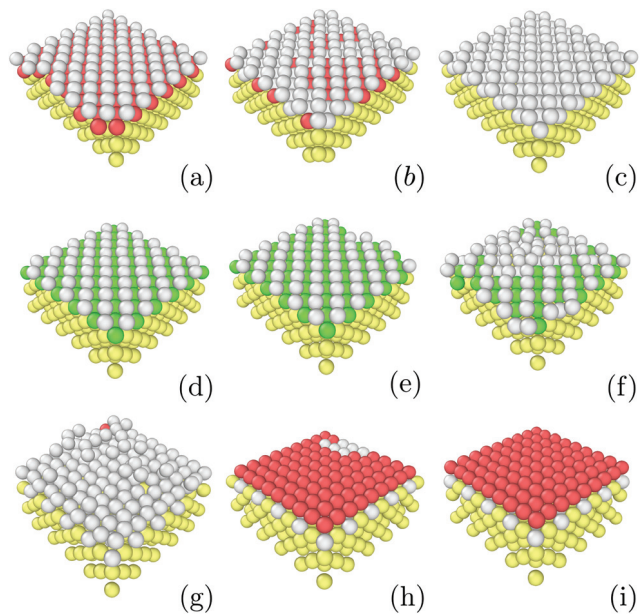

**Fig. 4** Images of structures emerging as the first layer obtained with $\varepsilon_{s/m} = 2.00$, $k_BT = 0.3$ and at different lattice strains: (a) $1.00a_s$, (b) $1.05a_s$, (c) $1.10a_s$, (d) $1.15a_s$, (e) $1.20a_s$, (f) $1.25a_s$, (g) $1.30a_s$, (h) $1.35a_s$ and (i) $1.40a_s$. Color designations are described in Fig. 1.

### 3.3 HCP to BCC transition

At $a_s = 1.20a_0$, the system transitions from being HCP rich to BCC rich when decreasing the size of the initial droplet at $k_BT = 0.5\varepsilon$ [see Fig. 3(c, f, i)]. In order to understand this striking result despite the fact that the first growing layer is made of BCC, we examine the temporal evolution leading to HCP. We note that during this temporal evolution no droplet shape evolution is observed [see Fig. 5(a–c)]. By fitting the HCP evolution with a hyperbolic tangent function, we extracted $\tau_{rise}$ as the time at which the HCP crystal phase undergoes the highest growth [see Fig. 6a]. In addition, BCC exhibits a peak just before the emergence of the HCP phase. $N_{BCC\rightarrow HCP}$, the total number of particles that have changed from BCC to HCP, sampled every $t_0$, is plotted in Fig. 6b. While large oscillations are found, after a transitory state, a plateau is observed. $N_{BCC\rightarrow HCP}$ is also fitted with the hyperbolic tangent function from which one obtains $\tau_{switch}$, the time when 95% of the plateau is reached. The agreement between $\tau_{rise}$ and $\tau_{switch}$ [see Fig. 6c] demonstrates that the emergence of HCP is triggered by the presence of BCC clusters turning into HCP. As shown in ditions where its presence builds an interface accommodating the FCC structures with the surrounding liquid.[54] Yet, to our knowledge, our work is a first observation of a cross-nucleation for the growth of HCP from BCC clusters where the BCC phase is not only present but is even predominant in the initial stages. This mechanism holds for $N = \{1; 2\} \times 10^3$ and $\varepsilon_{s/m} > 1.00\varepsilon$. Surprisingly, for larger droplets ($N = [1 \times 10^4; 5 \times 10^4]$), HCP remains the most preponderant structure even if FCC is more stable in bulk [see Fig. 7]. As evidenced by the identified two-step nucleation mechanism, the emergence of HCP is caused by the presence of the BCC first layers that constitute an appropriate template for HCP rather than FCC. Then, the HCP to FCC transformation can hardly occur within the simulation timescale because of the small free energy difference between the two structures and the very likely presence of an energy barrier. Finally, when the droplet is too small ($N = 5 \times 10^2$), the system remains BCC rich since the BCC clusters are too small to be converted into HCP [see Fig. 3(h and i)].

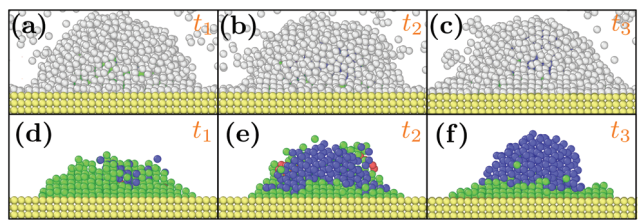

**Fig. 5** Snapshots of typical droplets during the freezing obtained for $N = 10^3$, $a_s = 1.20a_0$, $k_BT = 0.5\varepsilon$ and $\varepsilon_{s/m} = 2.00$. $t_1$, $t_2$ and $t_3$ are respectively equal to $50t_0$, $75t_0$ and $375t_0$. In (d, e, f), the liquid particles are hidden to better observe the droplet crystal structure. Color designations are described in Fig. 1.

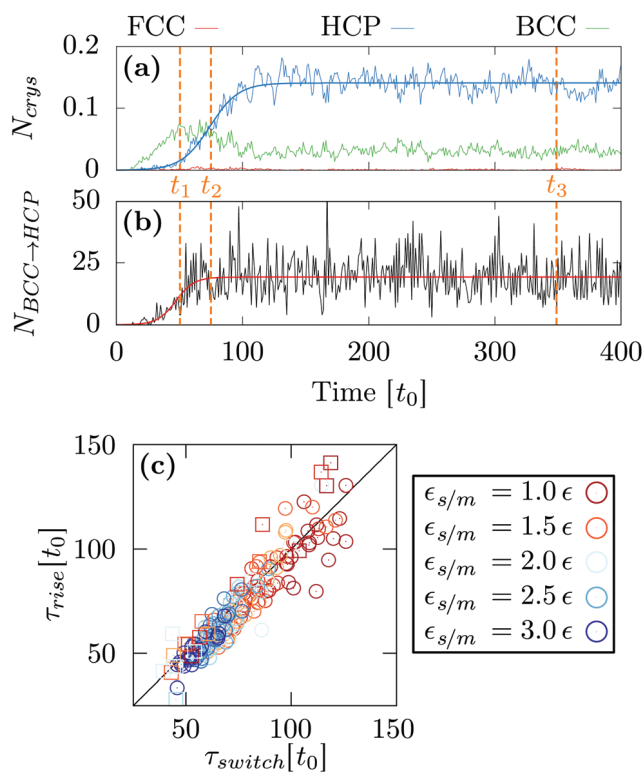

**Fig. 6** Temporal evolution of (a) the different crystal structures and (b) $N_{BCC\rightarrow HCP}$ obtained for $N = 10^3$, $a_s = 1.20a_0$, $k_BT = 0.5\varepsilon$ and $\varepsilon_{s/m} = 2.00$. (c) Comparison between $\tau_{rise}$ and $\tau_{switch}$. The circle and square represent respectively the results obtained with $N = 10^3$ and $N = 2 \times 10^5$. The solid line represents the $\tau_{rise} = \tau_{switch}$ curve.

Fig. 5(d–f), these clusters are located near the substrate. The mechanism for the cross-nucleation of these droplets is expected to be: (i) templated nucleation of BCC particles near the substrate as evidenced from Fig. 4, (ii) when the BCC cluster is large enough, transformation of BCC into HCP, and (iii) rapid growth of HCP. This result is consistent with the general mechanism of polymorph "cross nucleation" previously observed in the simulation of Lennard-Jones particles,[52–54] with Yukawa particles[55] and in experiments with more complex molecules.[56,57] In particular, the role of BCC in the nucleation process was already observed under bulk con-

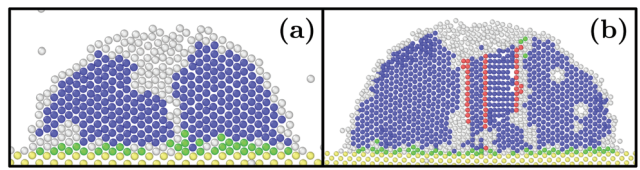

**Fig. 7** Slices of typical final states obtained with $\varepsilon_{s/m} = 2\varepsilon$, $k_B T = 0.3\varepsilon$, $a_s = 1.20 a_0$ and (a) $N = 1 \times 10^4$ and (b) $N = 5 \times 10^4$. Color designations are described in Fig. 1.

## 4 Conclusion

In summary, our study focused on the use of nanodroplet freezing on a crystalline substrate to engineer different structures. First, we showed that only the most hydrophilic substrates are able to trigger the crystallization. Then, increasing the substrate lattice parameter can drive the system into three distinct polymorphs: FCC, BCC and HCP. By probing the first layer to grow, we find that under intermediate strains, a BCC structure is indeed favorable which is easily understood as being epitaxial in origin. For small droplets, the stability of this BCC first layer is sufficient to obtain the growth of BCC across the entire droplet. For larger droplets, the volumetric part of the free energy becomes dominant and the conversion to HCP is thermodynamically favored. Since crystallization is obviously going to begin at the droplet–substrate interface, this explains why the final structure is made of a few layers of BCC at the substrate interface with HCP on top. Although the presence of HCP instead of FCC was previously observed in different contexts,[49,50,58,59] reordering into BCC was observed here for the first time using Lennard-Jones particles at ambient pressure.[60] Hence, while it could be experimentally difficult to change the substrate crystal structure, we demonstrated that even when the substrate crystalline structure remains FCC, a stretched lattice parameter is already able to stabilize different polymorphs, especially the exotic BCC phase. Therefore, these results suggest an original approach to engineer crystallites that would be unstable otherwise. Finally, for a particular lattice parameter ($a_s/a_0 = 1.2$), we showed that the emergence of HCP is triggered by the first step where BCC clusters are preponderant and located at the interface with the substrate. Afterwards, these BCC clusters transform into HCP at a particular time that coincides with the growth of HCP. In general, we evidenced a crucial role played by the BCC clusters which is surprising considering that the substrate crystalline structure is FCC. Since the studied model is rather elementary and the mechanisms are very general, these results could be validated by experiments involving high resolution transmission electron microscopy[9-11] and are thus expected to have impact on a broad variety of applications.

## Conflicts of interest

There are no conflicts to declare.

## Acknowledgements

The work of JL was funded by the European Union's Horizon 2020 research and innovation program within the AMECRYS project under grant agreement no. 712965. The work of JFL was funded by the European Space Agency under contract no. ESA AO-2004-070.